\documentclass[11pt]{article}

\newcommand{\BEAS}{\begin{eqnarray*}}
\newcommand{\EEAS}{\end{eqnarray*}}
\newcommand{\BEA}{\begin{eqnarray}}
\newcommand{\EEA}{\end{eqnarray}}
\newcommand{\BEQ}{\begin{equation}}
\newcommand{\EEQ}{\end{equation}}
\newcommand{\BIT}{\begin{itemize}}
\newcommand{\EIT}{\end{itemize}}
\newcommand{\BNUM}{\begin{enumerate}}
\newcommand{\ENUM}{\end{enumerate}}

\newcommand{\BA}{\begin{array}}
\newcommand{\EA}{\end{array}}

\newcommand{\ones}{\mathbf 1}

\newcommand{\reals}{{\mbox{\bf R}}}

\newcommand{\symm}{{\mbox{\bf S}}}  

\newcommand{\Tr}{\mathop{\bf Tr}}
\newcommand{\diag}{\mathop{\bf diag}}

\newcommand{\Expect}{\textstyle\mathop{\bf E{}}}

\newcommand{\QED}{~~\rule[-1pt]{6pt}{6pt}}

\newcommand{\vect}{\mathop{\bf vec}}

\newtheorem{remark}[theorem]{Remark}

\newcounter{exno}

{\begin{quote}}{\end{quote}}

\makeatletter
\long\def\@makecaption#1#2{
   \vskip 9pt 
   \begin{small}
   \setbox\@tempboxa\hbox{{\bf #1:} #2}
   \ifdim \wd\@tempboxa > 5.5in
        \begin{center}
        \begin{minipage}[t]{5.5in}
        \addtolength{\baselineskip}{-0.95pt}
        {\bf #1:} #2 \par
        \addtolength{\baselineskip}{0.95pt}
        \end{minipage}
        \end{center}
   \else 
    \hbox to\hsize{\hfil\box\@tempboxa\hfil}  
   \fi
   \end{small}\par
}
\makeatother

\newcounter{oursection}

\newcounter{lecture}

\usepackage{fullpage,times,amsmath}
\usepackage{amsfonts,amssymb}
\usepackage{psfrag,graphics,epsfig}

\begin{document}

\title{A Semidefinite Relaxation for\\ Air Traffic Flow Scheduling}

\author{Alexandre d'Aspremont\thanks{ORFE Department, Princeton University, Princeton, NJ 08544. \texttt{alexandre.daspremont@m4x.org}} \and Laurent El Ghaoui\thanks{EECS Department, U.C. Berkeley, Berkeley, CA 94720. \texttt{elghaoui@eecs.berkeley.edu}}}

\maketitle

\begin{abstract}
We first formulate the problem of optimally scheduling air traffic low with sector capacity constraints as a mixed integer linear program. We then use semidefinite relaxation techniques to form a convex relaxation of that problem. Finally, we present a randomization algorithm to further improve the quality of the solution. Because of the specific structure of the air traffic flow problem, the relaxation has a single semidefinite constraint of size $dn$ where $d$ is the maximum delay and $n$ the number of flights. 
\end{abstract}

\section{Introduction}
In this paper, given a schedule of flights and their routes we solve the problem of finding a new schedule that satisfies a list of sector capacity constraints and minimizes the total delay compared to the original scheduled. While the optimal routing problem with weather uncertainty and capacity constraints is essentially intractable, \cite{NEG05} show that robust optimal routing of a single aircraft under weather and traffic uncertainty can be solved efficiently as a robust Markov dynamic programming (MDP) problem. So the problem we solve here should be seen as a second phase: given the routes computed using MDP, our relaxation produces an optimal schedule, satisfying capacity constraints while minimizing the total delay. While this scheduling problem is significantly simpler than the global routing problem, it still combinatorial (and in fact NP-Hard), hence we need to formulate a relaxation to obtain approximate solutions efficiently. We can interpret this problem as particular case of job shop scheduling problem where a task corresponds to an aircraft occupying a sector at a certain time, with the aircraft routes correspond to lists of tasks that have to performed in sequence and where the sector capacities correspond to server capacities. In our case, a few key distinctions simplify the problem formulation. First, the different tasks are independent (one aircraft's route is not dependent on another). Second, for each aircraft, no time gap is allowed between two tasks (we assume that aircraft can't be held en-route).

Scheduling problems are notoriously hard combinatorial problems. Classic instances include the job-shop scheduling problem (see \cite{Coff76} for example) or the travelling salesman problem (see \cite{Rose77} for example). Semidefinite relaxations and randomization techniques have and excellent track record, which can be traced to \cite{Lova91}, \cite{Aliz95}, \cite{Goem95} and \cite{Polj95} among others. Recently, \cite{Fraz01} applied these techniques to aircraft conflict avoidance for free flight, by adjusting aircraft speed and bearing.

In this paper, we first formulate the air traffic flow scheduling problem as a mixed integer linear programming problem. We then apply the lifting procedure of \cite{Goem95} to formulate a semidefinite relaxation of the problem. Because of the structure of our problem here, we only have $dn$ binary variables, where $d$ is the maximum delay and $n$ the number of flights, which means that the semidefinite relaxation has one semidefinite constraint of size $dn+1$, allowing to scale better than classic scheduling problem relaxations. We then detail a randomization procedure based on that of \cite{Goem95} that uses the matrix solution of the semidefinite relaxation to further improve the quality of the solution. While \cite{Fraz01} focus on free flight and conflict avoidance, our main concern is on meeting sector capacity constraints, a key limiting factor in the European airspace, by adjusting delay at departure. This relative simplicity allows the algorithm to scale very well with the number of aircraft. Finally, because of its particular structure, we show that this relaxation is amenable to first-order methods for large-scale semidefinite optimization such as those detailed in \cite{helm00} and \cite{Nest04a} which are natural algorithms for solving large-scale problems for which a low precision is required.

The paper is organized as follows. The second section defines the main scheduling problem and formulates it as a mixed integer linear program. The third section derives the semidefinite relaxation of that problem. In the fourth section, we show how to exploit the result of this relaxation to further improve the solution using a randomization technique. Finally, in section five, we present some numerical results.

\section{Problem formulation}
\label{s:prob} Suppose we are given routes for $n$ aircraft flying across an airspace composed of $m$ sectors with capacities given by $C\in\reals^m$. We decompose a particular day into $T$ periods, so that a particular flight route starting at time $s$ can represented by a matrix $R^{(i,s)}\in\reals^{m\times T}$ such that: 
\[
\left\{\BA{ll}
R^{(i,s)}_{jt}=1 & \mbox{if aircraft $i$ is in sector $j$ at time $t$}\\
R^{(i,s)}_{jt}=0 & \mbox{if not.}
\EA\right.
\]
We can formulate the problem of minimizing total delay while satisfying capacity constraints as:
\BEQ\label{eq:milp}
\BA{ll}
\mbox{minimize} & \sum_{i=1}^n \sum_{j=0}^d x_{ij} j\\
\mbox{subject to} & \sum_{i=1}^n \sum_{j=0}^d x_{ij} R^{(i,s+j)} \leq C\\
& \sum_{j=0}^d x_{ij}=1\\
& x_{ij}\in\{0,1\}, \quad i=1,\ldots,n,~j=0,\ldots,d,
\EA
\EEQ
in the variable $x_{ij}\in\reals^{n \times (d+1)}$ where $d$ is the maximum delay (in units of time). Here, $x_{ij}=1$ means that aircraft $i$ will be delayed by $j$ units of time (because $x$ is a binary variable and $\sum_{j=0}^d x_{ij}=1$, there can only be a single nonzero $x_{ij}$ for each aircraft $i$). The first constraint  makes sure that sector capacity constraints are met and the objective is the sum of aircraft delays.

\section{Semidefinite Relaxation}
\label{s:sdp-relax} In this section, we apply the \emph{lifting} procedure detailed in \cite{Goem95} to form a semidefinite relaxation of problem (\ref{eq:milp}). We can rewrite (\ref{eq:milp}) as a non-convex quadratic program (QP):
\BEQ\label{eq:milp-qp}
\BA{ll}
\mbox{minimize} & \sum_{i=1}^n \sum_{j=0}^d x_{ij} j\\
\mbox{subject to} & \sum_{i=1}^n \sum_{j=0}^d x_{ij} R^{(i,s+j)} \leq C\\
& \sum_{j=0}^d x_{ij}=1\\
& x_{ij}^2-x_{ij}=0, \quad i=1,\ldots,n,~j=0,\ldots,d,
\EA
\EEQ
in the variable $x_{ij}\in\reals^{n \times (d+1)}$. We can rewrite this problem as:
\BEQ\label{eq:ncvx-qp}
\BA{ll}
\mbox{minimize} & \sum_{i=1}^n \sum_{j=0}^d x_{ij} j\\
\mbox{subject to} & \sum_{i=1}^n \sum_{j=0}^d x_{ij} R^{(i,s+j)} \leq C\\
& \sum_{j=0}^d x_{ij}=1, \quad i=1,\ldots,n\\
& X=\vect(x)\vect(x)^T\\
& \diag(X)=\vect(x)\\
\EA
\EEQ
in the variables $x_{ij}\in\reals^{n \times (d+1)}$ and $X\in\symm^{n(d+1)}$.  This is a non-convex quadratic program and is computationally hard. We now detail how to obtain a convex relaxation of problem (\ref{eq:ncvx-qp}) using Lagrangian duality.

\subsection{Lagrangian relaxation}
\label{ss:relax} Let us start from a general nonconvex quadratically constrained quadratic program (QCQP):
\BEQ\label{eq:non-cvx-qcqp}
\BA{ll}
\mbox{minimize}   & x^T P_0 x + q_0^T x + r_0 \\
\mbox{subject to} & x^T P_i x + q_i^T x + r_i \leq 0,
\quad i=1,\ldots,m,\\
\EA\EEQ
with variable $x\in\reals^n$, and parameters $P_i\in\symm^n$,
$q_i\in\reals^n$, and $r_i\in\reals$. We form the Lagrangian,
\[
L(x,\lambda)=x^T\left(P_0+\sum_{i=1}^{m}{\lambda_i P_i}\right)x +
\left(q_0+\sum_{i=1}^{m}{\lambda_i q_i}\right)^T x + r_0 +
\sum_{i=1}^{m}{\lambda_i r_i}.
\]
To find the dual function, we minimize over $x$, using the general formula (see example~4.5 in \cite{Boyd03}):
\[
\inf_{x\in\reals}x^T P x + q^T x + r=\left\{
\BA{l}
r-\frac{1}{4} q^T P^{\dag} q,\quad\mbox{if }P\succeq0\mbox{ and
}q\in\mathcal{R}(P)\\
-\infty,\quad\mbox{otherwise.}
\EA\right.
\]
The dual function is then:
\BEAS
g(\lambda) &=&\inf_{x\in\reals^n}L(x,\lambda)\\
&=&-\frac{1}{4}\left(q_0+\sum_{i=1}^{m}{\lambda_i
q_i}\right)^T \left(P_0+\sum_{i=1}^{m}{\lambda_i
P_i}\right)^{\dag} \left(q_0+\sum_{i=1}^{m}{\lambda_i q_i}\right)
+ \sum_{i=1}^{m}{\lambda_i r_i} + r_0.
\EEAS
We can form the dual of (\ref{eq:non-cvx-qcqp}), using Schur complements (cf.~\S A.5.5):
\BEQ
\label{eq:dual-qcqp}
\BA{ll}
\mbox{maximize}   & \gamma + \sum_{i=1}^{m}{\lambda_i r_i} + r_0\\
\mbox{subject to} & \left[\BA{cc}
\left(P_0+\sum_{i=1}^{m}{\lambda_i P_i}\right) &
\left(q_0+\sum_{i=1}^{m}{\lambda_i q_i}\right)/2\\
\left(q_0+\sum_{i=1}^{m}{\lambda_i q_i}\right)^T/2 & -\gamma\\
\EA\right] \succeq 0\\ & \lambda_i \geq 0,\quad i=1,\ldots,m,
\EA\EEQ
in the variable $\lambda\in\reals^m$. As the dual to (\ref{eq:non-cvx-qcqp}), this is a convex program, it is in fact a semidefinite program (SDP). This SDP is called the \emph{Lagrangian relaxation} of the nonconvex QCQP. It can be solved efficiently and gives a lower bound on the optimal value of the nonconvex QCQP. We form take the dual of program (\ref{eq:dual-qcqp}) (see \cite[\S5.9.2]{Boyd03}):
\BEQ\label{eq:sdp-relax}
\BA{ll}
\mbox{minimize}   & \Tr(X P_0) + q_0^T x + r_0 \\
\mbox{subject to} & \Tr(X P_i) + q_i^T x + r_i \leq 0,
\quad\mbox{i=1,\ldots,m,}\\
                  & \left[ \BA{cc}
                  X & x\\
                  x^T & 1 \EA \right]\succeq 0,\\
\EA
\EEQ
with variable $x\in\reals^n$, $X\in\symm^n$ and parameters $P_i\in\symm^n$,
$q_i\in\reals^n$, and $r_i\in\reals$. 

\subsection{A relaxation of the scheduling problem}
Using the results of the previous section, we can form the Lagrangian relaxation of (\ref{eq:ncvx-qp}) as:
\BEQ\label{eq:ncvx-qp-relax}
\BA{ll}
\mbox{minimize} & \sum_{i=1}^n \sum_{j=0}^d x_{ij} j\\
\mbox{subject to} & \sum_{i=1}^n \sum_{j=0}^d x_{ij} R^{(i,s+j)} \leq C\\
& \sum_{j=0}^d x_{ij}=1, \quad i=1,\ldots,n,\\
& \left[ \BA{cc}
        X & \vect(x)\\
        \vect(x)^T & 1 \EA \right]\succeq 0\\
& \diag(X)=\vect(x),\\
\EA
\EEQ
which is a semidefinite program in the variables $x_{ij}\in\reals^{n \times (d+1)}$ and $X\in\symm^{n(d+1)}$ and can be solved efficiently. The objective of this program is a \emph{lower bound} on the global solution.

\subsection{First-order methods}
An important structural property of problem (\ref{eq:ncvx-qp-relax}) is that, because $\diag(X)=\vect(x)$ and $\ones^Tx=n$ the matrix variable:
\[
\left[ \BA{cc}
        X & \vect(x)\\
        \vect(x)^T & 1 \EA 
\right]
\]
has constant trace equal to $n+1$. This means that the dual of (\ref{eq:ncvx-qp-relax}) is a maximum conic eigenvalue minimization problem for which efficient first-order methods such as the spectral bundle algorithm of \cite{helm00} and the optimal first-order method of \cite{Nest04a}. In Section \ref{s:num-example}, we present numerical experiments using the SBmethod code by \cite{helm00} on increasingly large randomly chosen problems. SBmethod solves large-scale dense instances for which it returns a solution $(x,X)$. It can also solve much larger problems by exploiting sparsity, in which case however it only returns the optimal $x$, which somewhat decreases the performance of the randomization methods detailed in the next section.

\section{Randomization}
\label{s:rand} The Lagrangian relaxation techniques developed in~\S\ref{ss:relax} provided lower bounds on the optimal value of the program in (\ref{eq:non-cvx-qcqp}), but did not however give any particular hint on how to compute good feasible points. The semidefinite relaxation in (\ref{eq:sdp-relax}) produces a positive semidefinite or covariance matrix together with the lower bound on the objective. In this section, we exploit this additional output to compute good approximate solutions with, in some cases, hard bounds on their suboptimality.

\subsection{Randomization}
In section \ref{ss:relax}, the original nonconvex QCQP:
\[
\BA{ll}
\mbox{minimize}   & x^T P_0 x + q_0^T x + r_0 \\
\mbox{subject to} & x^T P_i x + q_i^T x + r_i \leq 0,
\quad i=1,\ldots,m,\\
\EA
\]
was relaxed into:
\BEQ\label{eq:bidual-qcqp-random}
\BA{ll}
\mbox{minimize}   & \Tr(X P_0) + q_0^T x + r_0 \\
\mbox{subject to} & \Tr(X P_i) + q_i^T x + r_i \leq 0,
\quad i=1,\ldots,m,\\
                  & \left[ \BA{cc}
                  X   & x\\
                  x^T & 1 \EA \right]\succeq 0.\\
\EA
\EEQ
The last (Schur complement) constraint being equivalent to $X-xx^T \succeq 0$, if we suppose $x$ and $X$ are the solution to the relaxed program in (\ref{eq:bidual-qcqp-random}), then $X-xx^T$ is a covariance matrix. If we pick $x$ as a Gaussian variable with $x\sim\mathcal{N}(x,X-xx^T)$, $x$ will solve the nonconvex QCQP in (\ref{eq:non-cvx-qcqp}) ``on average'' over this distribution,
meaning:
\[
\BA{ll}
\mbox{minimize}   & \Expect (x^T P_0 x + q_0^T x + r_0) \\
\mbox{subject to} & \Expect (x^T P_i x + q_i^T x + r_i) \leq 0,
\quad i=1,\ldots,m,\\
\EA
\]
and a ``good'' feasible point can then be obtained by sampling $x$ a sufficient number of times, then simply keeping the best
feasible point. Of course the direct sampling technique above does not guarantee that a feasible point will be found. In particular, if the program includes an equality constraint, then this method will certainly fail. However, it is sometimes possible to directly project the random samples onto the feasible set. As we will see below, this is the case here 

\subsection{Randomized schedules}
Suppose we have solved
\[\BA{ll}
\mbox{minimize} & \sum_{i=1}^n \sum_{j=0}^d x_{ij} j\\
\mbox{subject to} & \sum_{i=1}^n \sum_{j=0}^d x_{ij} R^{(i,s+j)} \leq C\\
& \sum_{j=0}^d x_{ij}=1, \quad i=1,\ldots,n,\\
& \left[ \BA{cc}
        X & \vect(x)\\
        \vect(x)^T & 1 \EA \right]\succeq 0\\
& \diag(X)=\vect(x),\\
\EA\]
to get an optimal $x_{ij}\in\reals^{n \times (d+1)}$ and $X\in\symm^{n(d+1)}$. We can simply sample a Gaussian variable $u\sim\mathcal{N}(x,X-xx^T)$ and compute its projection $v$ on 
\[
\{0,1\}^{n \times (d+1)}\bigcap\left\{x\in\reals^{n \times (d+1)}:~\sum_{j=0}^d x_{ij}=1,~i=1,\ldots,n\right\}.
\]
We then compute the delay for each feasible random schedule $v$ and keep the best solution.

\section{Numerical Example}
\label{s:num-example} To fix ideas, let us begin by solving a very simple scheduling example. Suppose that there are only four sectors and two flights.
\begin{center}
\includegraphics[width=.15\textwidth]{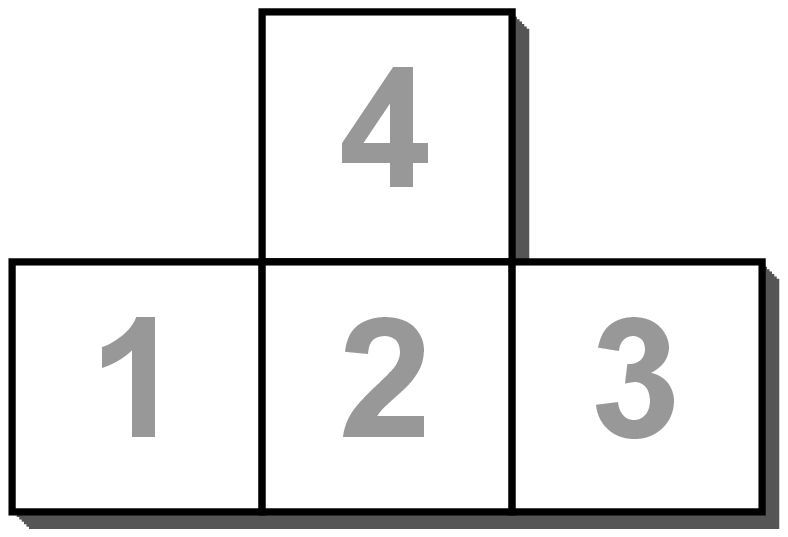}
\end{center}
Each sector has capacity one. In our format, a solution where both flights leave on time will be represented by $x=(1,0,1,0)$ and a solution where flight 1 is delayed by one unit of time will be written as $x=(0,1,1,0)$. Flight one starts from sector 1 and ends in sector 4, while flight 2 starts from sector 3 and also ends in sector 4. It takes each flight one unit of time to cross each sector and both flights are scheduled to depart at time 1 and conflict in sectors 2 and 3. The maximum delay is 1 unit of time. The problem variables are $x\in\reals^4$ and $X\in\symm^4$, the linear sector capacity constraints impose:
\[
\tiny{\left(\BA{cccc}
1 & 0 & 0 & 0 \\
0 & 0 & 0 & 0 \\
0 & 0 & 1 & 0 \\
0 & 0 & 0 & 0 \\
0 & 1 & 0 & 0 \\
1 & 0 & 1 & 0 \\
0 & 0 & 0 & 1 \\
0 & 0 & 0 & 0 \\
0 & 0 & 0 & 0 \\
0 & 1 & 0 & 1 \\
0 & 0 & 0 & 0 \\
1 & 0 & 1 & 0 \\
0 & 0 & 0 & 0 \\
0 & 0 & 0 & 0 \\
0 & 0 & 0 & 0 \\
0 & 1 & 0 & 1 \\
0 & 0 & 0 & 0 \\
0 & 0 & 0 & 0 \\
0 & 0 & 0 & 0 \\
0 & 0 & 0 & 0  
\EA \right)}x\leq \ones
\]
and the objective vector is given by $c=(0,1,0,1)$. We solve (\ref{eq:ncvx-qp-relax}) using SEDUMI by \cite{Stur99} to get the following solution:
\[
x=\small{\left(\BA{c}
1/2\\
1/2\\
1/2\\
1/2\\
\EA\right)}
\quad \mbox{and} \quad
X=\small{\left(\BA{cccc}
0.50 & 0.24 & 0.24 & 0.24 \\
0.24 & 0.50 & 0.24 & 0.24 \\
0.24 & 0.24 & 0.50 & 0.24 \\
0.24 & 0.24 & 0.24 & 0.50 \\  
\EA\right)}.
\]
Here, the solution is not binary valued. The reason for this is symmetry, the problem too simple and symmetric in flights one and two, so the solution is a mix of the two optimal solutions. This is easily fixed by adding a small random perturbation to the objective (implicitly breaking the tie by giving one aircraft priority over another). The solution is then:
\[
x=\small{\left(\BA{c}
1 \\
0 \\
0 \\
1 \\
\EA\right)}
\quad \mbox{and} \quad
X=\small{\left(\BA{cccc}
1 & 0 & 0 & 1 \\
0 & 0 & 0 & 0 \\
0 & 0 & 0 & 0 \\
1 & 0 & 0 & 1 \\ 
\EA\right)}.
\]
Here, the fact that $X=xx^T$ also shows that the relaxation is tight and that the solution $x$ is globally optimal.
\subsection{Computing times}
\begin{figure}[htbp]
\begin{center}
\begin{tabular}{cc}
\psfrag{n}[t][b]{n}
\psfrag{cpu}[b][t]{CPU time}
\includegraphics[width=0.45 \textwidth]{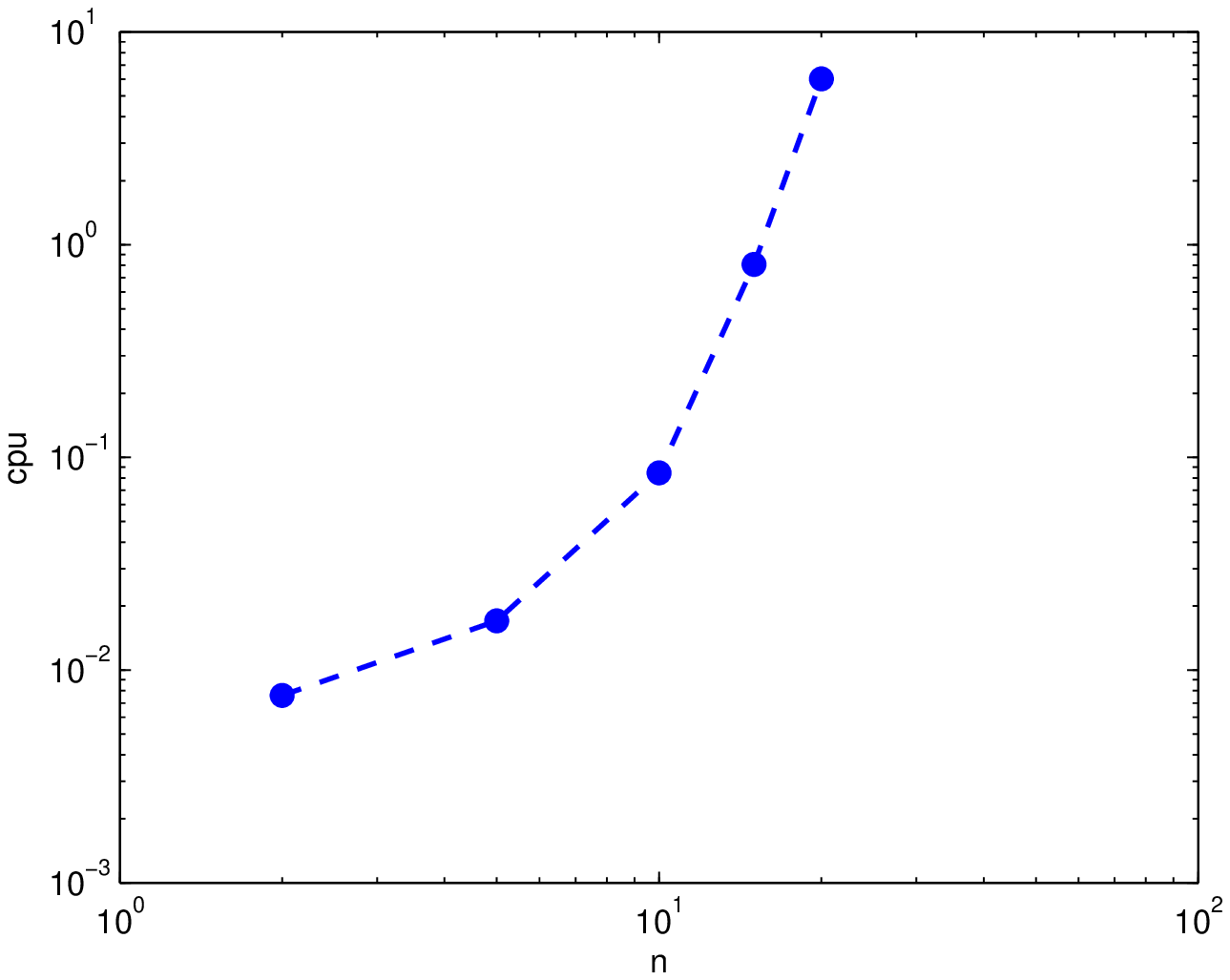}
&
\psfrag{n}[t][b]{n}
\psfrag{cpu}[b][t]{CPU time}
\includegraphics[width=0.45 \textwidth]{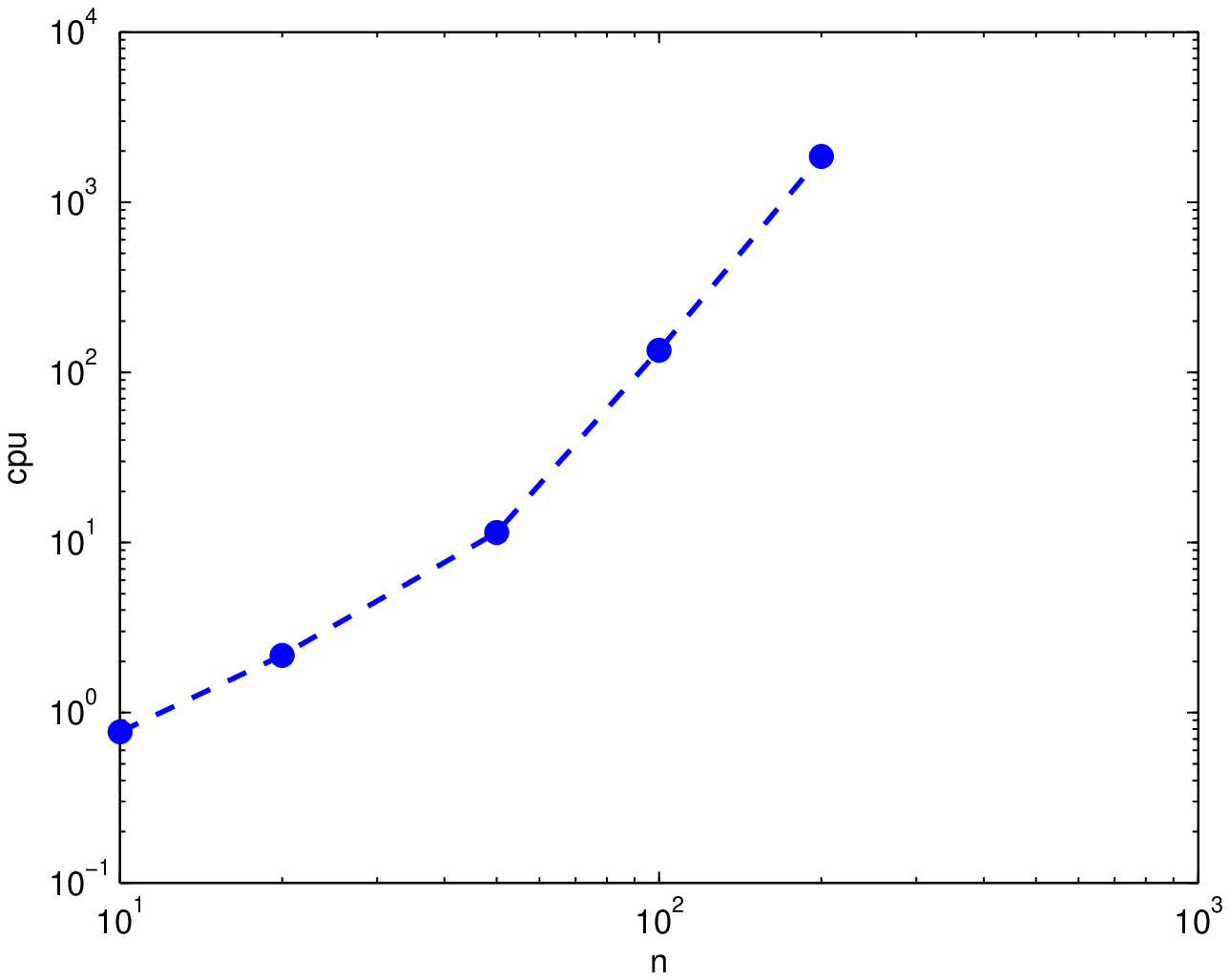}
\end{tabular}
\caption{Log-log plot of CPU time (in minutes) versus number of aircraft for randomly generated problems, \emph{left}: interior point methods, \emph{right}: spectral bundle method. \label{fig:cpu-vs-n}}
\end{center}
\end{figure}

Here, we generate random problems in an airspace with 50 sectors and a maximum delay of 2 units of time. In Figure \ref{fig:cpu-vs-n}, plot of CPU time (in minutes) versus number of aircraft for these problems.

With $n=20$ for example, the computing time is about 6 minutes, for a problem with $10^{18}$ possible schedules. While general-purpose interior point solvers such as SEDUMI by \cite{Stur99} can solve reasonably large problem instances. Using the fact that the semidefinite program in (\ref{eq:ncvx-qp-relax}) has constant trace,  more specialized first order methods (see \cite{helm00} or \cite{dasp04a} and \cite{dAsp05} for example) can be used to solve much larger problems.

\subsection{Randomization}
\begin{figure}[htbp]
\begin{center}
\psfrag{obj}[t][b]{Total delay}
\psfrag{freq}[b][t]{Samples}
\includegraphics[width=0.45 \textwidth]{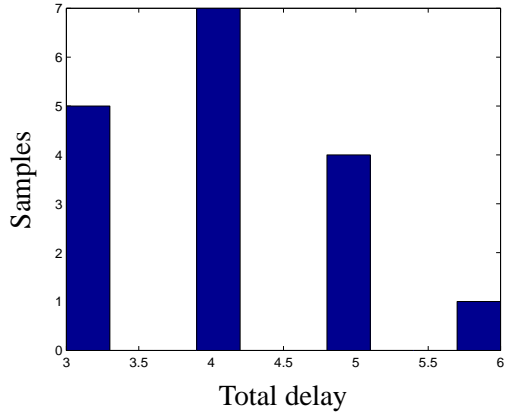}
\caption{Histogram of randomized objective values (Total delay). \label{fig:rand}}
\end{center}
\end{figure}

To illustrate the randomization method detailed in \S\ref{s:rand}, we first solve problem (\ref{eq:ncvx-qp-relax}) on a random problem with 5 sectors, 3 flights and a maximum delay of 4 units of time. The solution $x$ to (\ref{eq:ncvx-qp-relax})
has an objective value of 3, which gives a lower bound on the global optimum. We then generate 100 sample points from $u\sim\mathcal{N}(x,X-xx^T)$ and compute their projection $v$ on 
\[
\{0,1\}^{n \times (d+1)}\bigcap\left\{x\in\reals^{n \times (d+1)}:~\sum_{j=0}^d x_{ij}=1,~i=1,\ldots,n\right\}.
\]
For those sample schedules that meet the capacity constraints, we compute the total delay. Figure \ref{fig:rand} shows the distribution of these objective values.
We notice that the best delay found by randomization is equal to 3, which matches the lower bound produced by solving problem (\ref{eq:ncvx-qp-relax}). This shows that 3 is the globally optimum delay and that the corresponding randomized solution is optimal.

\section*{Acknowledgements}
The authors would like to acknowledge financial support from Eurocontrol grant C20083E/BM/05 and a gift from Google, Inc.

\bibliographystyle{alpha}
\small{\bibliography{Mainperso}}
\end{document}